\documentclass[10pt,conference]{IEEEtran}
\usepackage[printonlyused]{acronym}

\usepackage{amsmath,amssymb,amsfonts}
\usepackage{algorithmic}
\usepackage{graphicx}
\usepackage{array}
\usepackage{multirow}
\usepackage{booktabs}
\usepackage{textcomp}
\usepackage{xcolor}
\usepackage{mathtools}
\usepackage{placeins}
\usepackage{afterpage}
\usepackage{hyperref}
\usepackage{subfig}
\usepackage{multicol}
\usepackage{bm}
\usepackage{enumitem}
\usepackage{makecell}
\usepackage{colortbl}
\usepackage[size=small]{caption}
\usepackage{etoolbox}
\makeatletter
\patchcmd{\@makecaption}
  {\scshape}
  {}
  {}
  {}
\makeatletter
\patchcmd{\@makecaption}
  {\\}
  {.\ }
  {}
  {}
\makeatother
\def\tablename{TABLE}

\def\BibTeX{{\rm B\kern-.05em{\sc i\kern-.025em b}\kern-.08em
    T\kern-.1667em\lower.7ex\hbox{E}\kern-.125emX}}
{}

\makeatletter
\renewcommand{\fnum@figure}{Figure \thefigure}
\makeatother

\begin{document}

\title{Experimental Assessment of A Framework for In-body RF-backscattering Localization}
\vspace{-6mm}
\author{
\IEEEauthorblockN{Noa Jie Vives Zaguirre\IEEEauthorrefmark{1}, Oscar Lasierra\IEEEauthorrefmark{2}, Filip Lemic\IEEEauthorrefmark{2}\IEEEauthorrefmark{3}\textsuperscript{\textsection}, Gerard Calvo Bartra\IEEEauthorrefmark{2}, Pablo Jos\'e Galv\'an Calder\'on\IEEEauthorrefmark{2}, \\Gines Garcia-Aviles\IEEEauthorrefmark{2}\IEEEauthorrefmark{5}, Sergi Abadal\IEEEauthorrefmark{1}, Xavier Costa-P\'erez\IEEEauthorrefmark{2}\IEEEauthorrefmark{6}}
\vspace{2mm}
\IEEEauthorblockA{\IEEEauthorrefmark{1}Nanonetworking Center in Catalonia, Universitat Politècnica de Catalunya, Spain}
\IEEEauthorblockA{\IEEEauthorrefmark{2}AI-driven Systems Lab, i2CAT Foundation, Spain}
\IEEEauthorblockA{\IEEEauthorrefmark{3}Faculty of Electrical Engineering and Computing, University of Zagreb, Croatia}
\IEEEauthorblockA{\IEEEauthorrefmark{5}Intelligent Systems and Telematics Group, University of Murcia, Spain}
\IEEEauthorblockA{\IEEEauthorrefmark{6}NEC Laboratories Europe GmbH, Germany and ICREA, Spain\\
Email: filip.lemic@i2cat.net}
\vspace{-5mm}
}


\maketitle

\begingroup\renewcommand\thefootnote{\textsection}
\footnotetext{Corresponding Author.}
\endgroup

\begin{abstract}
Localization of in-body devices is beneficial for \ac{GI} diagnosis and targeted treatment. 
Traditional methods such as imaging and endoscopy are invasive and limited in resolution, highlighting the need for innovative alternatives. 
This study presents an experimental framework for \ac{RF}-backscatter-based in-body localization, inspired by the ReMix approach, and evaluates its performance in real-world conditions. 
The experimental setup includes an in-body backscatter device and various off-body antenna configurations to investigate harmonic generation and reception in air, chicken and pork tissues. The results indicate that optimal backscatter device positioning, antenna selection, and gain settings significantly impact performance, with denser biological tissues leading to greater attenuation. The study also highlights challenges such as external interference and plastic enclosures affecting propagation. The findings emphasize the importance of interference mitigation and refined propagation models to enhance performance.

\end{abstract}


\acrodef{GI}{Gastrointestinal}
\acrodef{RF}{Radio Frequency}
\acrodef{SNR}{Signal-to-Noise Ratio}
\acrodef{NLOS}{Non-Line-of-Sight}
\acrodef{GPS}{Global Positioning System}
\acrodef{PSD}{Power Spectral Density}
\acrodef{SISO}{Single-Input Single-Output}
\acrodef{MIMO}{Multi-Input Multi-Output}
\acrodef{USRP}{Universal Software Radio Peripheral}
\acrodef{ML}{Machine Learning}

\section{Introduction}

The prevalence of \acf{GI} diseases represents a global health challenge, affecting millions annually~\cite{wang2023global, riesgo2015specialization}. These conditions, ranging from acute issues such as gastroenteritis to chronic ones like inflammatory bowel disease, often present with nonspecific symptoms, such as abdominal pain and diarrhea~\cite{macaluso2012evaluation}. Such vague clinical presentations can lead to diagnostic errors and suboptimal treatment, underscoring the need for precise and efficient diagnostic tools.

\begin{figure}[t]
\centerline{\includegraphics[width=0.9\columnwidth]{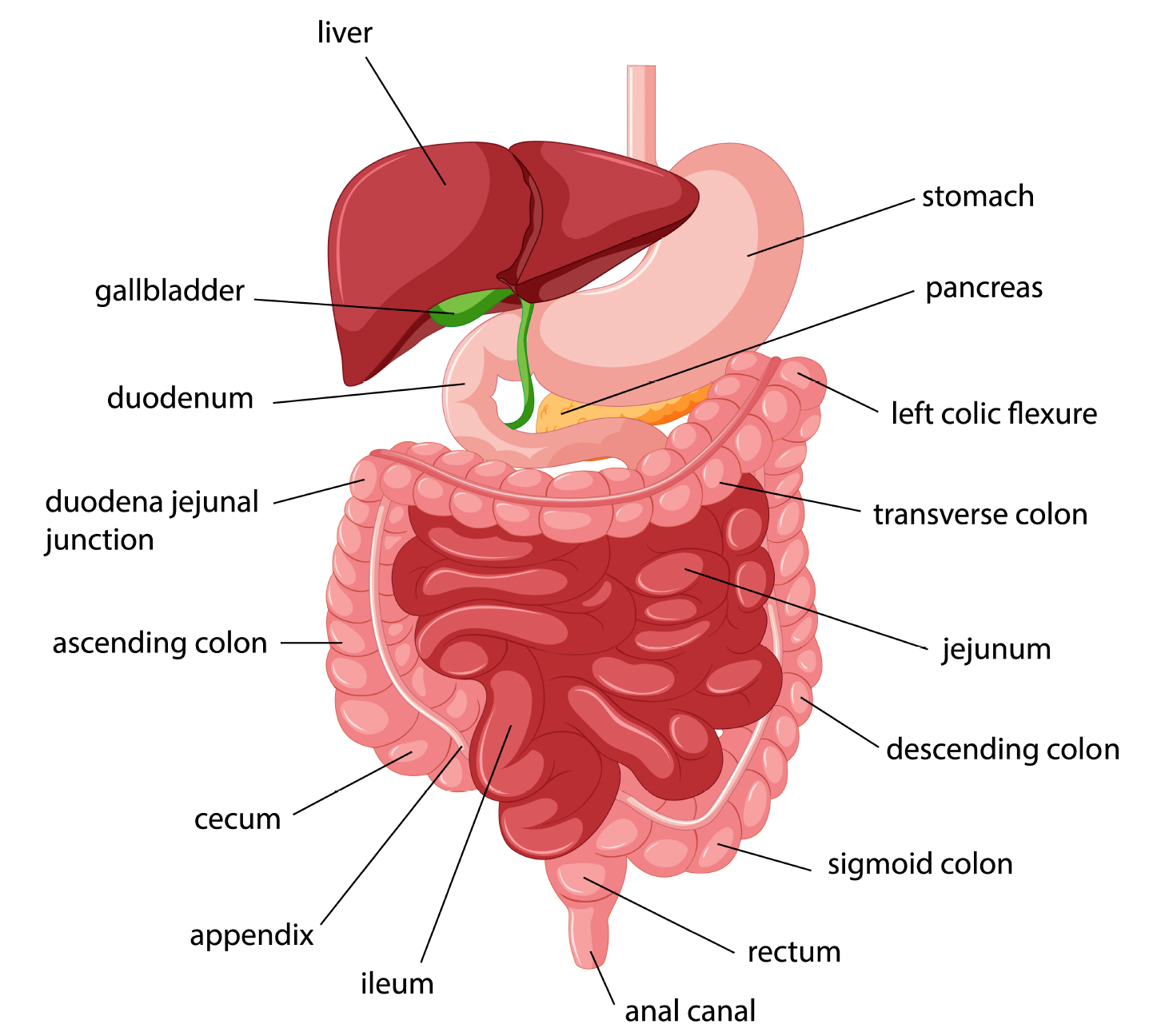}}
\caption{Representation of the GI tract}
\label{fig:gi_tract}
\vspace{-5mm}
\end{figure}

Accurate localization within the \ac{GI} tract is beneficial for early diagnosis and targeted treatments. Diagnostics in this field emphasize disease identification across specific \ac{GI} parts: the foregut (comprising the esophagus, stomach, liver, pancreas, and proximal duodenum), the midgut (i.e., distal duodenum, jejunum, ileum, cecum, appendix, ascending colon, and proximal two-thirds of transverse colon), and the hindgut (i.e., distal third of transverse colon, descending colon, sigmoid colon, rectum, and anal canal), as depicted in Figure~\ref{fig:gi_tract}. 
However, traditional diagnostic approaches, such as imaging and endoscopy, are costly, invasive, and often fail to provide sufficient diagnostic resolution~\cite{moore2003advantages,than2012review,vasisht2018body}.

Ingestible capsules have emerged as a promising alternative for diagnosing \ac{GI} diseases. These capsules are painless, sedation-free, and capable of localization and tracking, but their low granularity limits diagnostic utility~\cite{cao2024robotic}. Despite their potential, commercially available solutions are also limited by their reliance on internal batteries, which occupy valuable space and increase capsule size. Furthermore, prototype solutions lack the localization accuracy necessary for practical application, fail to account for complex signal propagation in the human body, and rely on static or simplified models that do not consider dynamic interactions, such as attenuation, multipath, and reflections~\cite{cao2024robotic}.

Toward addressing these limitations, \acf{RF}-backscattering has been proposed as a promising in-body localization approach. Notably, the ReMix approach~\cite{vasisht2018body} leverages a diode-antenna system that uses backscattering to mitigate interference from skin reflections. Specifically, ReMix employs sophisticated non-linear processing to mix input frequencies and generate harmonics, effectively filtering out unwanted reflections. Its layered linear path models account for signal propagation through different tissue types, such as skin, fat, and muscle. This approach has demonstrated exceptional localization accuracy, achieving an average error of 1.4~cm in chicken tissue and 1.27~cm in human phantoms~\cite{vasisht2018body}.

However, while ReMix demonstrates the potential of \ac{RF} backscattering localization using out-of-band harmonics, its evaluation was conducted in highly controlled \ac{RF} environments, such as shielded chambers. However, real-world applications require these algorithms to perform effectively in complex and uncontrolled \ac{RF} environments. The gap between performance in controlled settings and practical deployments highlights the importance of validating these algorithms under real-world conditions.

To address these challenges, this paper introduces an experimental framework for evaluating in-body \ac{RF}-backscattering localization. 
The framework includes a setup designed to collect raw data from an in-body backscatter device in realistic and uncontrolled \ac{RF}  environments.
By conducting experiments ``in the wild,'' this framework not only facilitates objective performance evaluation but also accounts for the practical challenges faced during real-world deployments. While this approach inherently results in lower observed performance compared to controlled experiments, it provides valuable insights into the robustness and reliability of the algorithms such as ReMix.

The key contribution of this work is the development of a framework tailored to \ac{RF}-backscattering in-body localization, enabling reproducible raw data collection in real-world \ac{RF} environments. The framework implements a modular approach for supporting the evaluation of various localization algorithms, guaranteeing fair and standardized performance benchmarks. Finally, we validate the framework using a backscatter device embedded in biological tissue, showcasing its utility in assessing algorithm performance under realistic conditions.


\section{RF-based In-body Localization}

\subsection{State of the Art}

\ac{GI} capsule endoscopy has traditionally relied on image recognition for capsule localization~\cite{dove2014analysis}. These capsules capture images in real time, and image recognition algorithms identify anatomical features such as folds or blood vessels, enabling region-based localization. However, this method has significant limitations in accuracy and reliability, and relies heavily on post-processing, and delays diagnosis and treatment.

\begin{figure}[t]
\centerline{\includegraphics[width=0.93\columnwidth]{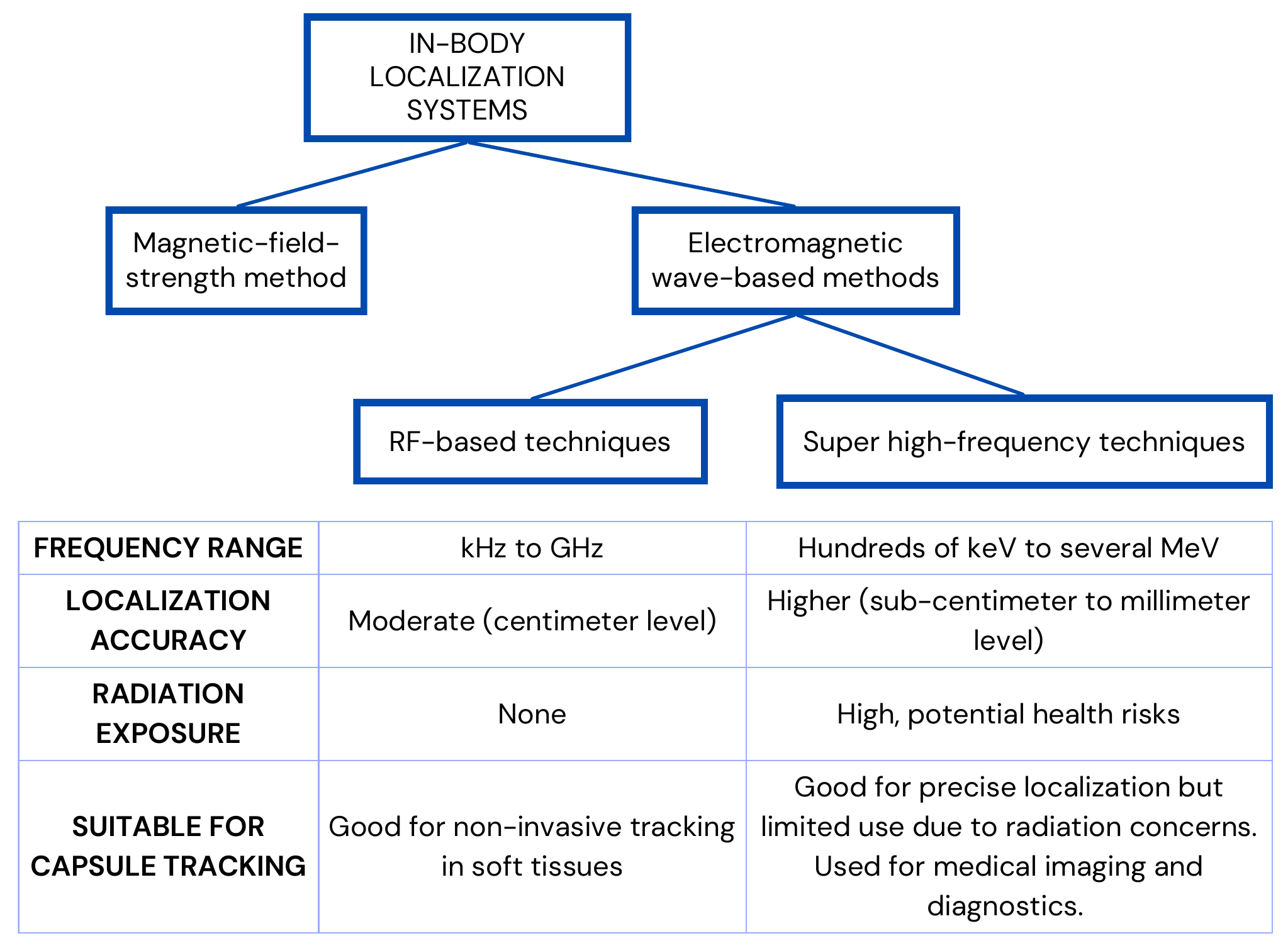}}
\vspace{-1mm}
\caption{Overview of in-body localization methods}
\label{fig:sota}
\vspace{-4mm}
\end{figure}

Two primary approaches emerged toward addressing these limitations (cf., Figure~\ref{fig:sota}): magnetic-field-strength and electromagnetic wave-based methods~\cite{than2012review,vasisht2018body}. 
Magnetic methods leverage the low absorption of magnetic fields by human tissues, making them safer compared to \ac{RF}-based techniques. They provide high accuracy and granularity, as discussed in~\cite{vasisht2018body}. However, these systems face challenges, including calibration complexity, limited generalizability across patients, and the requirement for specialized infrastructure~\cite{vasisht2018body,bianchi2019localization}. 

Electromagnetic methods, particularly those relying on \ac{RF}, have emerged as a promising alternative. \ac{RF}-based localization methods operate within the kHz to GHz range and are advantageous due to their deep penetration in human tissue, low-cost hardware, and adaptability to various diagnostic applications~\cite{bianchi2019localization,paret2009rfid}. In contrast to high-frequency methods like X-rays and gamma rays, which carry safety risks at higher doses, \ac{RF}-based techniques enable real-time and automatic capsule tracking while maintaining safety standards. 

\subsection{Challenges}

Despite its potential, in-body \ac{RF} localization in the GI track presents unique challenges not encountered in more traditional localization systems. The primary issue arises from the interaction of \ac{RF} signals with biological tissues. As \ac{RF} signals propagate through the body, they encounter layers with varying dielectric properties, such as muscle, bone, and fat. Each layer absorbs signals differently, causing significant attenuation and unpredictable propagation. Interfaces between tissues create impedance mismatches, leading to signal reflections, scattering, and highly multipath propagation with self-interference, negatively affecting localization accuracy~\cite{vasisht2018body}.

Miniaturization and power efficiency impose additional constraints. Capsules must be ingestible, necessitating the downsizing of antennas and batteries. However, smaller antennas reduce efficiency, weakening signal strength and complicating signal detection. Passive power sources can help reduce capsule size but require energy-efficient designs to sustain \ac{RF} transmission for localization~\cite{vasisht2018body}. Additionally, the directionality of signal escape poses a challenge, as only signals near the surface normal can exit the body, while wider-angle signals are internally reflected~\cite{vasisht2018body}. To address this, multiple external anchors are often required, increasing system complexity.

Achieving millimeter-level accuracy is another challenge. Unlike traditional systems such as \ac{GPS}, which target meter-level accuracy in outdoor environments, in-body localization must contend with the variability of human anatomy. Differences in tissue composition, layer thickness, and individual anatomy significantly affect signal propagation and necessitate patient-specific system adaptations.
Finally, in-body systems must adhere to strict safety and biocompatibility standards. The materials used for constructing antennas and other components must be safe for prolonged exposure to the human body, further limiting design options.


\section{Framework for Evaluation of In-body RF-based Localization Approaches}
\label{sec:model}

RF-backscattering-based localization has emerged as a promising approach for tracking in-body devices, particularly for applications such as ingestible \ac{GI} capsules. While prior studies have demonstrated its feasibility, their evaluations have often been conducted in controlled environments, leaving open questions about its real-world performance.

\begin{figure}[t]
\centerline{\includegraphics[width=\columnwidth]{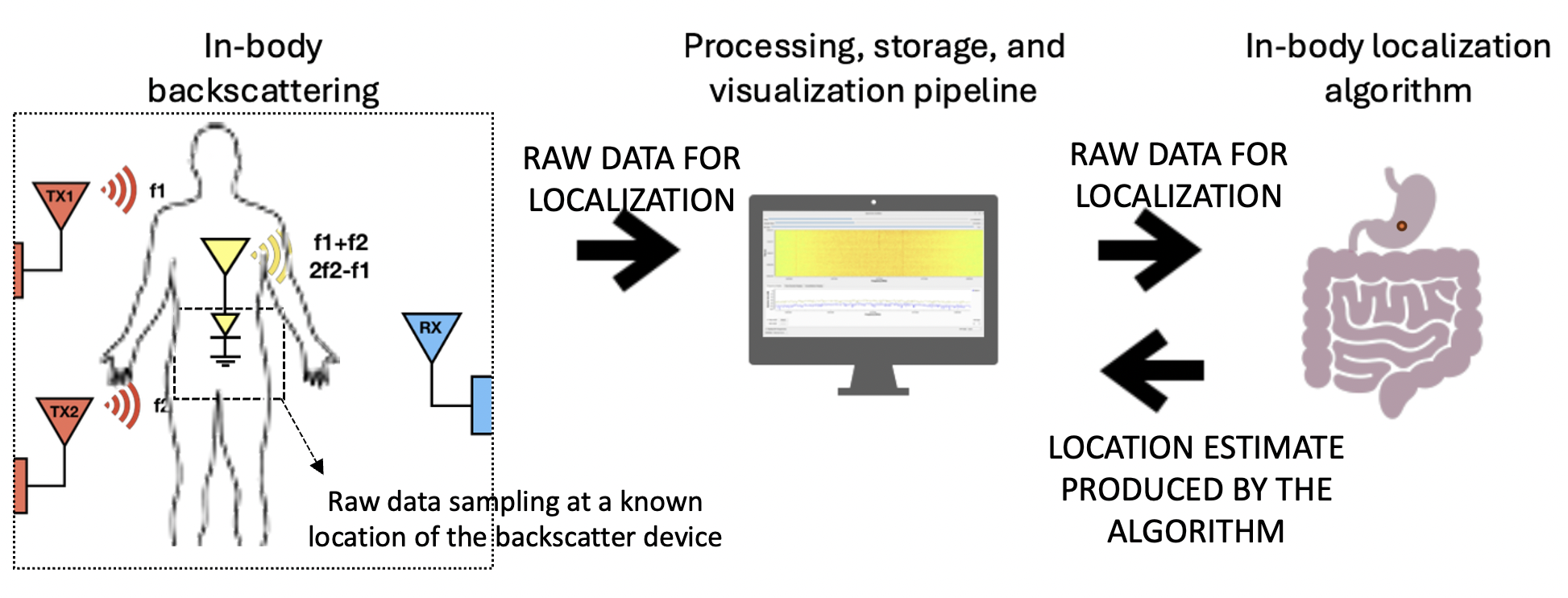}}
\caption{Design of the localization framework}
\label{fig:design}
\vspace{-4mm}
\end{figure}

This section describes the framework to assess the practicality of RF backscattering-based in-body localization under more realistic conditions.
The framework is envisaged to allows us to investigate the impact of environmental factors and system parameters, providing insights into key challenges such as signal propagation, interference, and hardware configurations.
The framework  integrates a flexible hardware setup, a carefully designed backscatter device, and robust signal processing tools to enable the evaluation of in-body RF-based localization approaches (cf., Figure~\ref{fig:design}). The modular design of the framework allows for iterative improvements and adaptation to different experimental needs, paving the way for future advancements in RF-based localization systems.

\subsection{Hardware Setup}

The framework is compatible with both \ac{SISO} and \ac{MIMO} setups to test and validate the localization framework. These setups provide flexibility for system evaluation, beginning with simpler configurations and progressing to more complex ones.

The SISO setup utilizes one transmitting and one receiving antenna, each connected to a \ac{USRP} device. The transmitting antenna generates a sine wave at a fixed carrier frequency, with adjustable parameters such as sample rate, amplitude, and transmission gain. 
The receiving antenna is aligned at a fixed distance from the transmitter to capture the transmitted signal. The received signals are stored for post-processing, enabling the evaluation of signal quality through metrics such as \ac{PSD} and phase response. The simplicity of the SISO configuration facilitates initial testing, allowing developers to address potential hardware or signal processing issues before transitioning to more complex setups.

The MIMO setup builds on the SISO configuration for creating a multi-antenna system. Each transmitter operates at a unique frequency to facilitate multi-frequency localization experiments. For example, two antennas may transmit signals at different frequencies while the receiving antennas capture and process these signals simultaneously.

Synchronization between the transmitting and receiving antennas is managed through the \ac{USRP} devices, ensuring that multi-frequency signals are accurately captured. This configuration allows for more comprehensive evaluations of system performance in scenarios where multiple frequencies are required to achieve high localization accuracy.

In both configurations, the transmitting antennas are programmed using a custom script to emit carrier signals with user-defined parameters, while the receiving antennas use tools like \textit{inspectrum}\footnote{Available at: \url{https://github.com/miek/inspectrum}.} to store the received signals in a complex binary format. This enables detailed analysis of captured data and ensures flexibility in adjusting the transmission and reception parameters for different experimental needs.

\subsection{Backscatter Device}

The backscatter device plays a crucial role in generating harmonics and enabling RF-based localization experiments. In this work, it consists of the following components:

\begin{itemize}
    \item \textbf{Antenna}: A dipole antenna (i.e., Taoglas PC30 flat patch antenna) designed for operation within the desired frequency range. The antenna is chosen for its compatibility with the transmitted signals and its ability to maintain stable performance under varying conditions.
    \item \textbf{Diode}: A Schottky detector diode (i.e., SMS7621-006LF) that mixes input signals received by the antenna to generate harmonics. These harmonics are subsequently backscattered to the receiving antennas for analysis.
\end{itemize}

The assembly process involves directly soldering the antenna wires to the diode. The ground wire of the antenna is connected to the cathode of the diode, while the signal wire is connected to the anode. This configuration ensures that the diode can effectively generate harmonics by combining the signals received at its input frequencies.

The current backscatter device is designed for operation in-air, with its dimensions and gain optimized for such use cases. However, in-body applications require miniaturization of the antenna to fit within an ingestible device. Emerging technologies in miniaturized antennas for in-body applications have demonstrated that antennas comparable in size to a grain of rice can be successfully implemented for \ac{RF}-based localization systems~\cite{lodato2014numerical}.

\subsection{GNU Radio for Signal Visualization}

GNU Radio~\cite{blossom2004gnu} provides a modular platform for processing and visualizing the signals captured by the receiving antennas. Its flexibility allows for real-time signal monitoring and post-experiment analysis, supporting the following tasks:

\begin{itemize}
    \item \textbf{Frequency response analysis}: Visualizing the relative gain of the received signal across a range of frequencies to identify distortions or losses during transmission.
    \item \textbf{Spectrogram analysis}: Generating time-frequency plots to monitor the stability of transmitted signals and identify harmonics generated by the backscatter device.
\end{itemize}

The processing pipeline is configured to save captured signals in a format compatible with tools like \textit{inspectrum}, enabling  analysis of \ac{PSD}, phase, and other key metrics. This raw data forms the foundation for streamlined performance evaluation of in-body localization algorithms, facilitating their objective back-to-back comparison and benchmarking.


\section{Considered In-body Localization Solution}

We use the ReMix algorithm~\cite{vasisht2018body} as the system under test. The algorithm determines the position of an in-body device by processing backscattered signals received by multiple antennas and applying optimization techniques to estimate the device's location. It operates through the following pipeline:

\begin{itemize}
\item \textbf{Phase Extraction and distance calculation:} The algorithm extracts the phase components of the received signals to calculate the effective distances between the backscatter device and each receiving antenna. These distances incorporate the effects of signal propagation through tissues, accounting for variations in attenuation and reflection caused by different tissue layers.

\item \textbf{Optimization for localization:} To estimate the device position, the algorithm minimizes the error between the calculated and measured distances. It applies local non-linear minimization for estimating signal phases and propagation. Adam optimizer is used to iteratively refine the estimated position and other system parameters.
\end{itemize}

The ReMix algorithm is highly configurable, allowing customization of key input parameters:
\begin{itemize}
\item \textbf{Signal frequencies:} Default carrier frequencies are 830~MHz and 870~MHz.
\item \textbf{Relative permittivity values:} The algorithm incorporates relative permittivity values for muscle and fat, with defaults set to 54.81 and 5.447, respectively~\cite{andreuccetti2012internet}.
\item \textbf{Antenna Positions:} Receiving antennas' positions are defined as 3D coordinates, enabling flexible configurations.
\end{itemize}

The primary output of the algorithm is the estimated 3D position of the backscatter device, calculated relative to the reference grid (cf., Figure~\ref{fig:setup}). This estimated position is then compared to the ground truth location of the device to calculate the localization error, defined as the Euclidean distance between the estimated and ground truth locations, providing a direct measure of localization accuracy.


\section{Evaluation Methodology}

This section describes the evaluation methodology used to validate the implemented prototype and the localization algorithm. The methodology is divided into two parts: evaluation of the hardware prototype, including the transmission and reception (Tx/Rx) chains, backscatter device, and signal processing pipeline; and evaluation of the localization algorithm using harmonic signals generated during the experiments.

\subsection{Evaluation of the Implemented Prototype}

\begin{figure}
\vspace{-3mm}
\centering
\subfloat[\centering In-body setup]{{\includegraphics[width=0.16\linewidth]{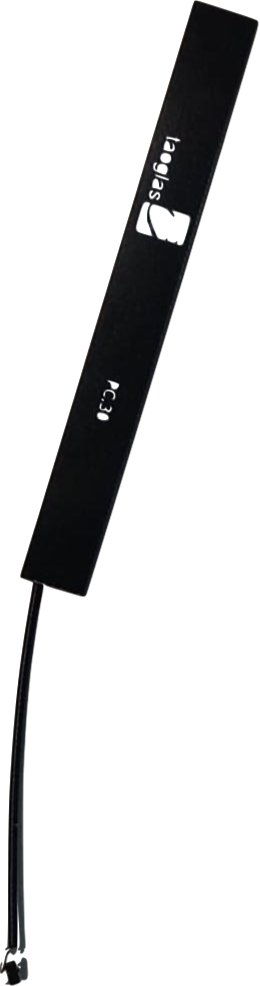} }}%
\subfloat[\centering Tx/Rx and evaluation setup]{{\includegraphics[width=0.83\linewidth]{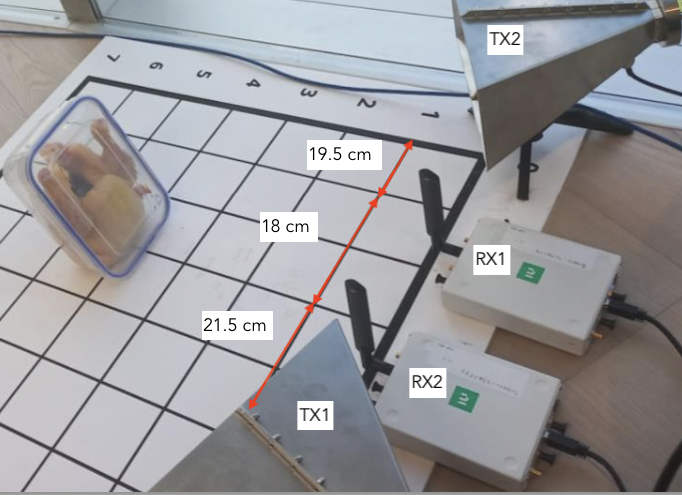} }}%
\vspace{-3mm}
\caption{Hardware setup}
\label{fig:setup}
\vspace{-4mm}
\end{figure}

A 70 × 70~cm grid served as the reference for positioning the antennas and the backscatter device, as shown in Figure~\ref{fig:setup}. Transmitting antennas were connected to \ac{USRP} B210 devices and aligned along the x-axis of the grid, while receiving antennas were also positioned along the x-axis at calculated distances to minimize interference and ensure optimal reception. 
Two configurations were tested during the experiments. In \textbf{Configuration 1}, horn antennas were used for transmission and omnidirectional antennas for reception. In \textbf{Configuration 2}, horn antennas were used for both transmission and reception. The in-body setup was enclosed and positioned with the antenna vertically oriented, elevated 7.4~cm above the ground.

The prototype was tested in three scenarios: \textbf{in-air}, where the backscatter device was positioned in front of the receiving antennas; \textbf{in-chicken}, where the device was placed inside chicken tissue (approximately 16 × 22 × 10~cm); and \textbf{in-pork}, where the device was embedded in pork (approximately 8 × 12 × 1~cm). The transmission gain was fixed at 70~dB for both carrier frequencies during all tests. Omnidirectional antennas were positioned 1~cm above the ground, while horn antennas were elevated to the height of 9.5~cm.

Signals at \(f_1 = 830\) and\(f_2 = 870~\text{MHz}\) were transmitted, and their harmonics (\(2f_1 - f_2 = 910\) and \(f_{1} + f_{2} = 1700~\text{MHz}\)) were captured by the receivers. To determine antenna and backscatter device positions that maximizes the relative gain due to the backscatter device, the device was incrementally moved along the y-axis of the grid from 0 to 70~cm in 10~cm steps. At each position, the \ac{PSD} of the harmonics was compared in scenarios with and without the backscatter device, verifying its contribution to harmonic generation.

\subsection{Evaluation of the Localization Algorithm}

Once the prototype's performance was validated, the ReMix algorithm was tested using harmonic signals recorded during the chicken and pork experiments. The recorded signals served as input to the localization pipeline. The algorithm processed the harmonic signals to estimate the 3D position of the backscatter device.

Key parameters were configured for each experiment. The carrier frequencies were input into the pipeline, along with the relative permittivity values for muscle, fat, and air. The positions of the transmitting and receiving antennas were defined as shown in Table~1, ensuring alignment with the experimental setup. The algorithm used the phase information from the harmonic signals to compute the effective distances between the backscatter device and the antennas. These effective distances were then used to estimate the 3D position of the device through an optimization process.


\section{Evaluation Results}

This section presents the experimental results obtained from evaluating the proposed \ac{RF}-backscattering localization framework under different conditions, including in-air, in-chicken, and in-pork scenarios. The analysis focuses on optimal backscatter device placement, signal gain variations, and localization accuracy.

\subsection{Optimal Placement of the Backscatter Device and Antennas}

The optimal placement of the backscatter device was determined based on the relative gain and the ability of the receiving antennas to capture the generated harmonics. Table~\ref{tab:optimal_placement} summarizes the most effective placement of the backscatter device and antennas under different conditions. The placement of the Tx and Rx antennas varied across configurations. Configuration 1 used horn antennas for transmission and omnidirectional antennas for reception, whereas Configuration 2 employed horn antennas for both transmission and reception. Across all experimental conditions, the optimal placement of the backscatter device ranged from approximately 25 to 38~cm along the x-axis and 10.5 to 22~cm along the y-axis.

In Configuration 1.B, where the receiving antennas were repositioned to improve signal reception, the localization performance improved significantly in the in-pork scenario. The adjustments made in Configuration 1.B optimized the relative gain for the harmonics. The data presented in Table~\ref{tab:optimal_placement} shows that small adjustments in the placement of the antennas and backscatter device significantly influence the strength of received harmonics.

\begin{table}[t]
    \centering
    \caption{Optimal (x,y,z) [cm] position of the backscatter device for indicated antennas positions. N/A: no position found with a higher value of PSD due to the backscatter device}
    \small
    \resizebox{\columnwidth}{!}{
    \begin{tabular}{l c c c c c}
        \hline
        \textbf{\makecell{Experiment/\\Configuration}} & \textbf{TX2} & \textbf{RX1} & \textbf{RX2} & \textbf{TX1} & \textbf{\makecell{Backscatter\\Device}} \\
        \hline
        AIR/C1 & (-1.5, 0, 9.5) & (18, 0, 8) & (36, 0, 8) & (57.5, 0, 9.5) & (30, 18, 11.1) \\
        AIR/C2 & (-5, 0, 9.5) & (20, 0, 9.5) & (45, 0, 9.5) & (70, 0, 9.5) & (25, 18, 11.1) \\
        \hline
        CHICKEN/C1 & (-1.5, 0, 9.5) & (18, 0, 8) & (36, 0, 8) & (57.5, 0, 9.5) & (30, 22, 11) \\
        CHICKEN/C2 & (-5, 0, 9.5) & (20, 0, 9.5) & (45, 0, 9.5) & (70, 0, 9.5) & N/A \\
        \hline
        PORK/C1 & (-1.5, 0, 9.5) & (18, 0, 8) & (36, 0, 8) & (57.5, 0, 9.5) & N/A \\
        PORK/C1.B & (-1.5, 0, 9.5) & (15, 0, 8) & (33, 0, 8) & (57.5, 0, 9.5) & (25, 10.5, 6) \\
        PORK/C2 & (-5, 0, 9.5) & (20, 0, 9.5) & (45, 0, 9.5) & (70, 0, 9.5) & (29, 15.5, 12) \\
        \hline
    \end{tabular}}
    \label{tab:optimal_placement}
    \vspace{-3mm}
\end{table}

\subsection{Relative Gain Analysis}

The performance of the backscatter device was evaluated based on the relative gain measurements at 910 MHz and 1700 MHz under different experimental conditions. Table~\ref{tab:relative_gain_distances} presents the relative gain differences observed when the backscatter device was positioned at various distances from the antennas. In the in-air experiments, harmonics were detected with high signal strength across all configurations, but when the device was placed inside biological tissue, the received signal strength decreased significantly due to absorption and scattering effects. The attenuation was more pronounced at 1700 compared to 910~MHz, highlighting the challenges associated with using higher frequencies for in-body localization.

At shorter distances, specifically below 10~cm, a reduction in relative gain was observed for certain harmonics, leading to degraded localization performance. Conversely, at longer distances, the harmonics remained detectable, although their power levels decreased with increasing distance. The results shown in Table~\ref{tab:relative_gain_distances} further demonstrate that at distances beyond approximately 60~cm the relative gain of the backscatter device diminished to the point where it became indistinguishable from background noise.


\begin{table}[t]
    \centering
    \caption{Relative gain [dB] due to the backscatter device}
    \resizebox{\columnwidth}{!}{
    \begin{tabular}{l c c c c c}
        \hline
        \textbf{Experiment} & \textbf{Backscatter} & \multicolumn{2}{c}{\textbf{RX1 [MHz]}} & \multicolumn{2}{c}{\textbf{RX2 [MHz]}} \\
        \cline{3-6}
        & \textbf{Position} & \textbf{910} & \textbf{1700} & \textbf{910} & \textbf{1700} \\
        \hline
        AIR (10-60~cm) & (25, 68, 11.1) & -58.71 & -73.57 & -67.88 & -78.92 \\
        Rx gain = 30~dB   & No backscatter & -66.76 & -77.90 & -83.48 & -79.76 \\
        & Difference & \cellcolor{green!20}8.05 & \cellcolor{green!20}4.33 & \cellcolor{green!20}15.6 & \cellcolor{green!20}0.84 \\
        \hline
        PORK (10-60~cm) & (36, 60.5, 12) & -74.01 & -74.01 & -83.82 & -82.59 \\
        Rx gain = 20~dB    & No backscatter & -75.24 & -74.62 & -85.05 & -83.82 \\
        & Difference  & \cellcolor{green!20}1.23 & \cellcolor{green!20}0.61 & \cellcolor{green!20}1.23 & \cellcolor{green!20}1.23 \\
        \hline
        AIR ($<$10~cm) & (30, 0, 9.5) & -46.25 & -66.25 & -56.25 & -70.00 \\
        Rx gain = 40~dB & No backscatter & -48.19 & -69.38 & -51.88 & -63.75 \\
        & Difference & \cellcolor{green!20}1.94 & \cellcolor{green!20}3.13 & \cellcolor{red!20}-4.37 & \cellcolor{red!20}-6.25 \\
        \hline
        PORK ($<$10~cm) & (25, 10.5, 9.5) & -77.6 & -74.62 & -85.21 & -85.21 \\
        Rx gain = 20~dB & No backscatter & -75.24 & -74.62 & -85.05 & -83.82 \\
        & Difference & \cellcolor{red!20}-2.36 & \cellcolor{red!20}0.00 & \cellcolor{red!20}-0.16 & \cellcolor{red!20}-1.39 \\
        \hline
    \end{tabular}}
    \vspace{-3mm}
    \label{tab:relative_gain_distances}
\end{table}

\subsection{Impact of Tissue Type on Signal Propagation}

\begin{figure}
\centering
\subfloat[\centering Chicken (configuration 1)]{{\includegraphics[width=0.92\linewidth]{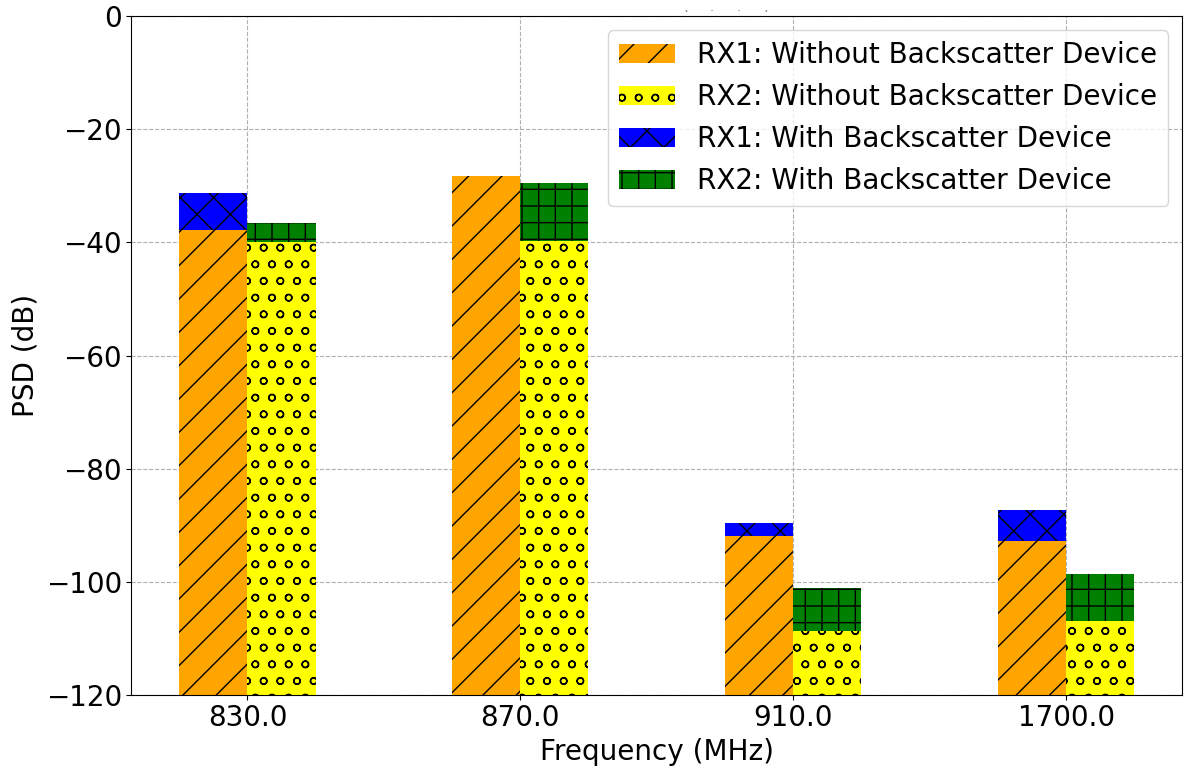} }}%
\\
\subfloat[\centering Pork (configuration 1)]{{\includegraphics[width=0.92\linewidth]{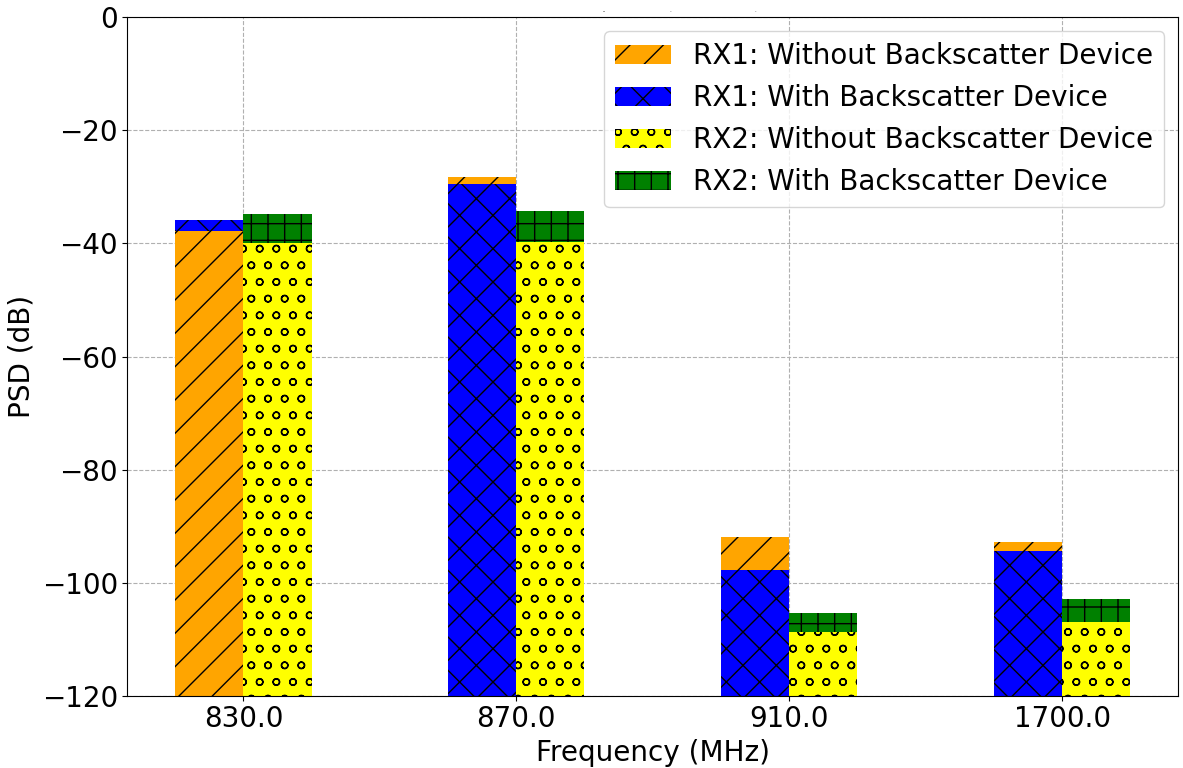} }}%
\caption{Impact of tissue type on PSD}
\label{fig:psd_chicken_pork}
\vspace{-5.5mm}
\end{figure}

The type of biological tissue used in the experiments had a significant impact on the reception of harmonic signals. Figure~\ref{fig:psd_chicken_pork} compares the \ac{PSD} values recorded for the in-chicken and in-pork experiments, illustrating the differences in signal attenuation. Pork tissue was found to attenuate harmonic signals more than chicken tissue, particularly at 1700~MHz, where the received power was consistently lower. The structural differences between the two tissues likely contributed to this effect. Chicken tissue contained air pockets and openings that allowed harmonics to escape, reducing the amount of absorption and scattering. In contrast, pork tissue was denser, leading to greater attenuation and a lower received power.

The experimental validation in pork tissue demonstrated that signal propagation characteristics differed significantly from those observed in in-air and in-chicken scenarios. As shown in Table~\ref{tab:relative_gain_distances}, the choice of gain settings played a critical role in maximizing the detectability of harmonics in biological tissue. A gain setting of 20~dB provided the best trade-off between signal strength and noise suppression, particularly for in-body applications where higher gain settings amplified both the desired signal and background interference.

\subsection{Impact of Antenna Configuration on Localization Accuracy}

\begin{figure}
\centering
\subfloat[\centering Horn antennas (configuration 1)]{{\includegraphics[width=0.92\linewidth]{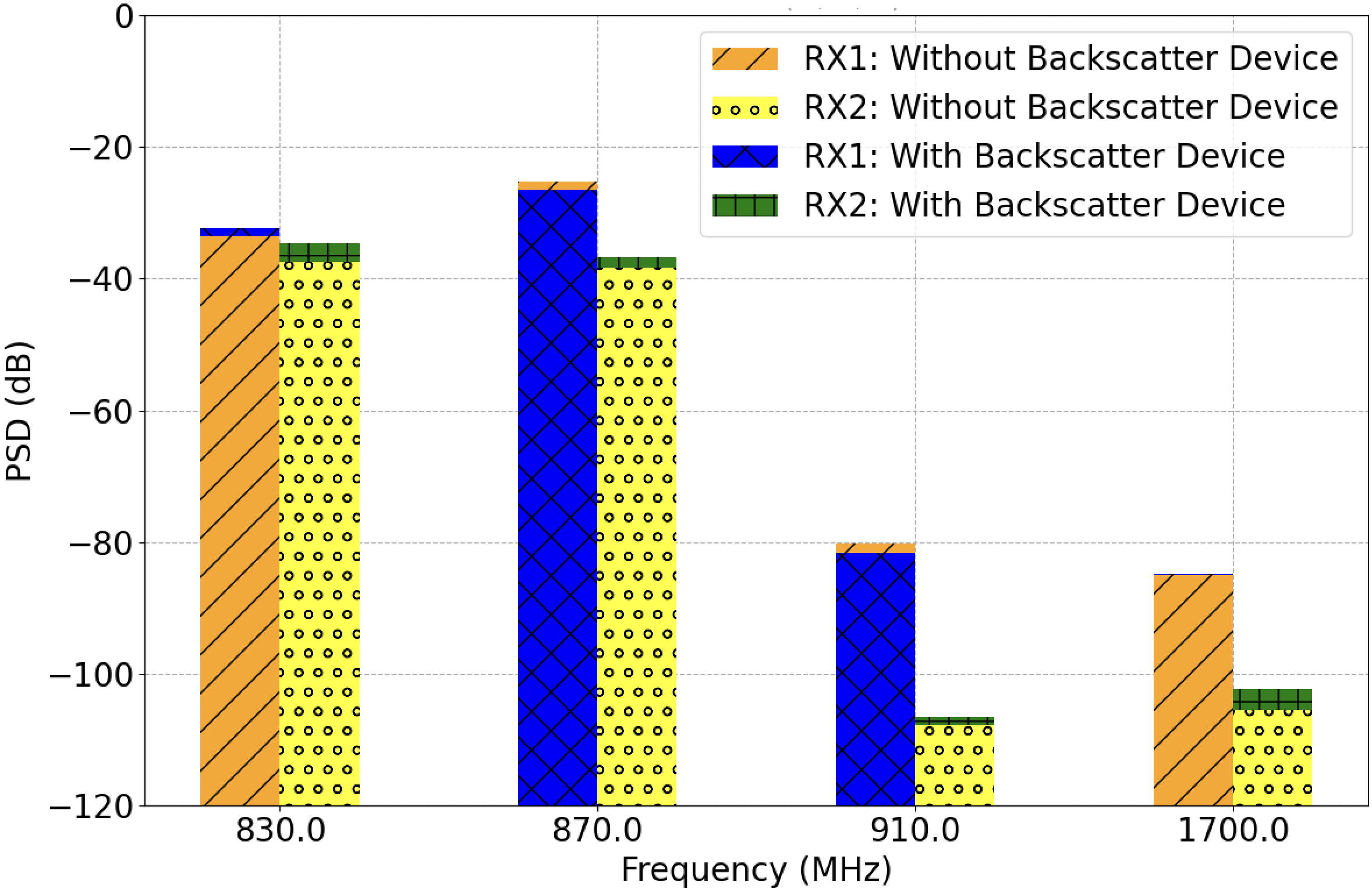} }}%
\\
\subfloat[\centering Omnidirectional antennas (configuration 2)]{{\includegraphics[width=0.92\linewidth]{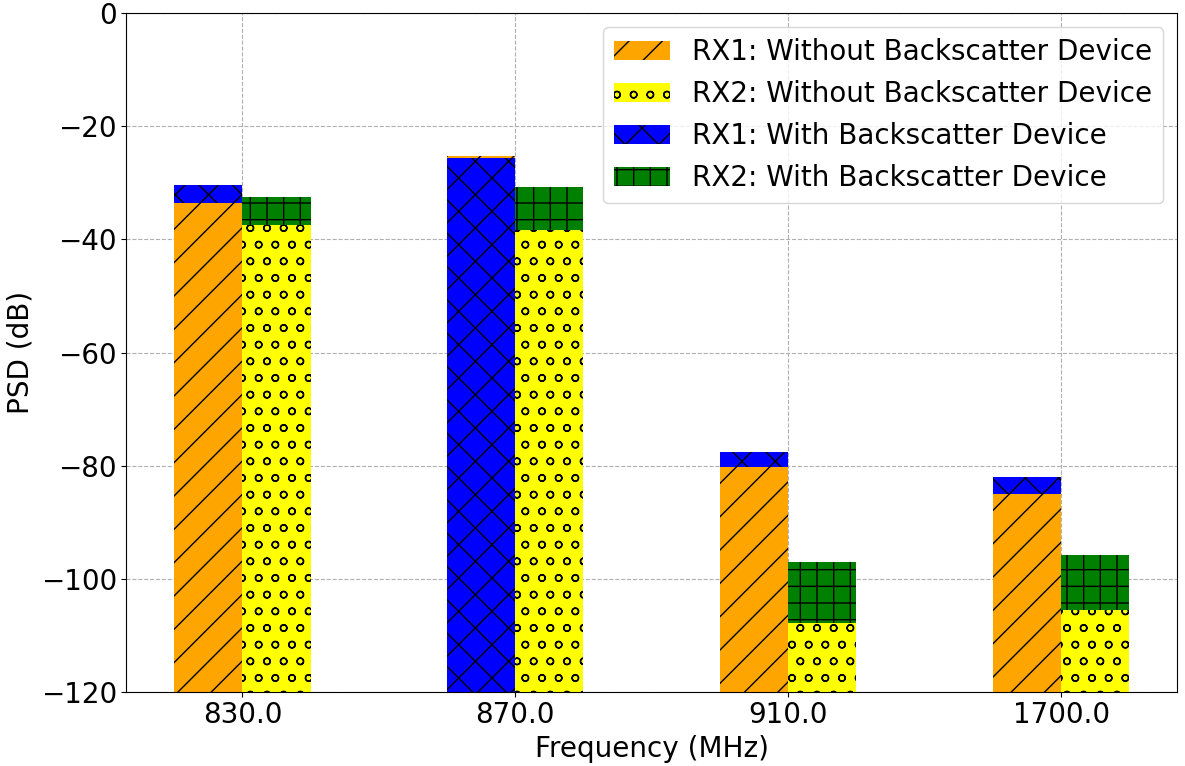} }}
\caption{Impact of antenna directionality on PSD}
\label{fig:antenna}
\vspace{-5.5mm}
\end{figure}

The choice of antenna type and placement had a direct effect on the accuracy of the localization system. Omnidirectional antennas captured signals more effectively in biological tissue. 
The performance of horn antennas in biological tissues is lower then their omnidrectional counterparts due to their directional nature, which limited the ability to capture backscattered harmonics affected by multipath propagation. 
Figure~\ref{fig:antenna} supports this observation by showing that omnidirectional antennas performed better in in-chicken experiments, where signal reflections within the tissue played a crucial role in harmonic reception.

\subsection{Localization Accuracy and Algorithm Performance}

\begin{figure}[t]
\centerline{\includegraphics[width=\columnwidth]{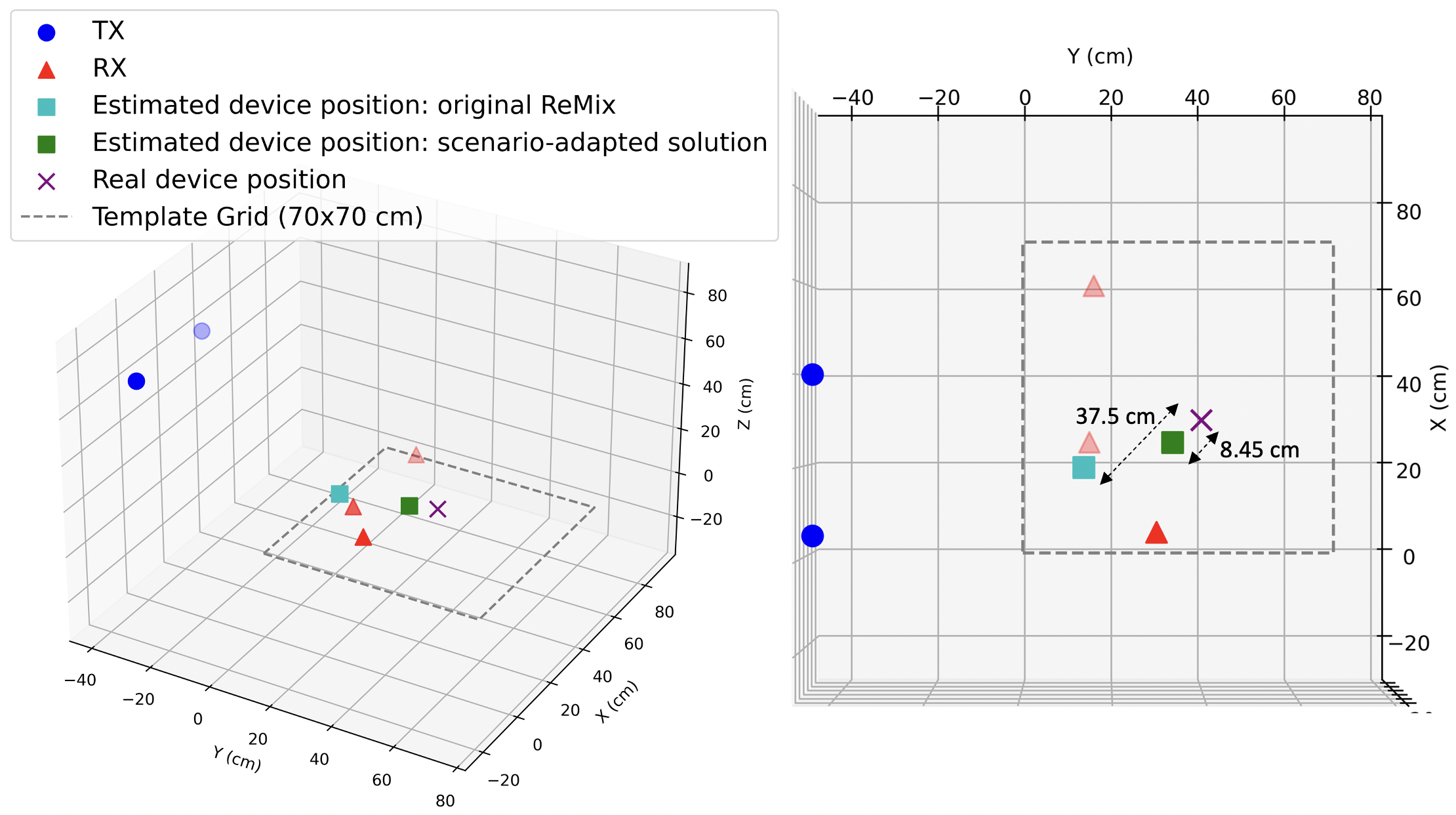}}
\caption{In-pork validation with three receiving antennas: 3D representation of the antenna placements, along with the actual and estimated positions of the backscatter device}
\label{fig:localization_3rx}
\vspace{-5mm}
\end{figure}

ReMix was used for obtaining the estimated positions from the recorded harmonic signals, followed by comparing them to corresponding ground truth locations. 
The average localization error across all configurations and scenarios ranged from 8.45 to 37.5~cm. Figure~\ref{fig:localization_3rx} shows the estimated and actual positions of the backscatter device in an example experimental configuration, illustrating the error distribution in the x, y, and z dimensions. The best estimated locations were obtained for the in-pork scenario with three receiving antennas.
The estimated locations are depicted for our original implementation of ReMix, as well as for the one optimized for the considered scenario, which yielded the best localization error of 8.45~cm.

\section{Discussion}
\label{sec:conclusions}

A key finding of this study is the importance of optimizing backscatter device positioning and antenna configuration to enhance signal reception and harmonic generation. The results indicate that specific placement ranges and gain settings significantly impact performance, with different environmental conditions requiring tailored adjustments. Proper positioning within a defined spatial range and tuning of transmission parameters improve the effectiveness of the system, suggesting the need for environment-specific calibration.

The material properties of biological tissues were another critical factor influencing signal attenuation and reception. The results indicate that denser tissues attenuate harmonics more than less dense tissues, primarily due to differences in permittivity, thickness, and structure. Additionally, experimental conditions such as plastic enclosures may have impacted signal behavior, highlighting the need to incorporate plastic layers into future propagation models for increased accuracy.

Antenna type also played a pivotal role in reception performance. Horn antennas proved advantageous for detecting signals from distant sources but required highly optimized positioning, while omnidirectional antennas required less optimization for optimal reception. Given the complexity of in-body environments, an adaptive or hybrid antenna configuration may offer a viable path forward.

The localization algorithm achieved an average accuracy of 8.45~cm when utilizing three receiving antennas. However, reducing the number of antennas to two resulted in localization errors between 15.71 and 37.5~cm, with particularly large ones observed along the y-axis. This degradation highlights the importance of additional antennas for robust in-body localization.
The main limitations encountered were raw measurement distortions due to external interference and plastic enclosures. 
Consequently, these raw data distortions negatively affected the localization algorithm's performance, mirroring challenges commonly faced in traditional \ac{RF}-based indoor localization systems~\cite{lemic2015experimental,behboodi2015interference}. 
These findings underscore the necessity of interference mitigation techniques and refined synchronization to enhance the system's reliability in practical deployments.

\section{Conclusion and Future Work}

This study provides an experimental validation of an RF-backscattering-based in-body localization framework, extending previous research beyond controlled environments. The results demonstrate that, while backscatter devices can improve harmonic generation and facilitate localization, practical deployment requires careful tuning of several parameters, including antenna configuration, gain settings, and tissue-specific adjustments.
The framework successfully evaluates signal propagation, interference effects, and localization accuracy under diverse conditions. Despite performance variations across materials, the study validates the feasibility of non-invasive, RF-based localization for ingestible devices and highlights key areas for further development.

Future improvements should focus on interference mitigation, exploring noise cancellation techniques such as adaptive filtering, frequency hopping, or \ac{ML}-based suppression to improve signal clarity and localization precision.
Enhanced synchronization should be investigated to reduce phase misalignment errors, improving overall system accuracy.
Refined propagation models should be developed to account for the impact of plastic enclosures on RF behavior, improving localization accuracy and real-world reliability.
Additional experimental validation should be conducted by testing different antenna placements, frequency selections, and assessing localization performance in human phantoms to approximate real tissue conditions. The system should also be tested for localization and tracking of a moving device within a 3D model to more realistically simulate real-world applications.
By addressing these challenges, RF-based in-body localization can move closer to real-world deployment in medical diagnostics, improving the precision and reliability of ingestible and implantable medical devices.

\section*{Acknowledgments}
This work has been supported by the CERCA Programme of the Generalitat de Catalunya.

\bibliographystyle{ieeetr}  
\bibliography{biblio}


\end{document}